\documentclass[12pt,preprint]{aastex}

\def\etal{{et~al.\/}}
\def\tetal{{et~al.\ }}
\def\ie{{i.e.,}\ }

\def\refit{{\null}}
\def\kms{{km~s$^{-1}$}}

\begin{document} 

\title{A Measurement of the Contamination in [O~III] $\lambda 5007$ Surveys
of Intracluster Stars and the Surface Density of $z=3.13$ Ly$\alpha$ 
Galaxies}

\author{Robin Ciardullo\altaffilmark{1}}
\affil{Department of Astronomy and Astrophysics, The Pennsylvania State 
University \\ 525 Davey Lab, University Park, PA 16802}
\email{rbc@astro.psu.edu}

\author{John J. Feldmeier\altaffilmark{1}}
\affil{Department of Astronomy, Case Western Reserve University 10900 Euclid
Ave. \\ Cleveland, OH 44106-1712}
\email{johnf@eor.astr.cwru.edu}

\author{Kara Krelove}
\affil{Department of Astronomy and Astrophysics, The Pennsylvania State 
University \\ 525 Davey Lab, University Park, PA 16802} 
\email{kara@astro.psu.edu}

\author{George H. Jacoby}
\affil{WIYN\altaffilmark{2} Observatory, P.O. Box 26732, Tucson, AZ 85726}
\email{jacoby@noao.edu}

\and

\author{Caryl Gronwall}
\affil{Department of Physics and Astronomy, The Johns Hopkins University
\\ 3400 N. Charles Street, Baltimore, MD 21218}
\email{caryl@pha.jhu.edu}

\altaffiltext{1} {Visiting Astronomer, Cerro Tololo Inter-American 
Observatory, National Optical Astronomical Observatory, 
which is operated by the Association of Universities for Research
in Astronomy, Inc., under cooperative agreement with the National Science 
Foundation.}

\altaffiltext{2}{The WIYN Observatory is a joint facility of the University of 
Wisconsin-Madison, Indiana University, Yale University, and the National 
Optical Astronomy Observatory.}

\begin{abstract}
We present two pieces of evidence supporting the hypothesis that the bright
[O~III] $\lambda 5007$ sources found in Virgo's intracluster space are
intracluster planetary nebulae, rather than [O~II] $\lambda 3727$ 
galaxies at $z \sim 0.35$ or Ly$\alpha$ sources at $z \sim 3.13$.
First, we confirm the nature of five ``overluminous'' [O~III] sources that are 
postulated to lie in front of M87: by examining the composite spectrum of 
these objects, we show that the weaker [O~III] line at $\lambda 4959$ is 
definitely present at a strength $\sim 1/3$ that of [O~III] $\lambda 5007$.  
The ratio demonstrates that, at most, only one of the five objects is a 
background galaxy.  We then estimate the surface density of background 
emission-line objects by conducting a wide-field (0.13 deg$^2$) search at 
$\lambda 5019$ for faint emission line sources in a ``blank field'' located 
well away from any galaxy or cluster.  We show that the density of blank field 
emission-line sources is significantly lower than the density of sources 
detected between the galaxies of Virgo, but in good agreement with the
density of Ly$\alpha$ galaxies found by \citet{hcm}.  The implication is 
that background galaxies only account for $\sim 20\%$ of the planetary nebula 
candidates in Virgo's intracluster fields.

\end{abstract}


\section{Introduction}
A very powerful method for probing the history of galaxy clusters is through
the observation and analysis of intracluster light 
\citep[cf.][]{d84,mil83,mer84,mqb00}.  Recently, this field has been 
revolutionized by on-band/off-band [O~III] $\lambda 5007$ surveys for 
planetary nebulae (PNe) located between the galaxies of Fornax and Virgo.  
By searching for emission line objects on CCD frames and adopting reasonable 
values for the ratio of PNe to main sequence stars, \citet{tw97}, 
\citet{men97}, \citet{ciar98} and \citet{fcj98} all concluded that a 
substantial fraction ($\gtrsim 20\%$) of the stellar mass of these clusters
resides outside of any galaxy.

This result, however, is by no means secure.  By comparing star counts in 
a Virgo intracluster field with counts performed on the {\sl Hubble Deep
Field,} \citet{ftv98} concluded that only 
$\sim 11\%$ of Virgo's stellar mass belongs to a diffuse intergalactic
component.  In addition, the interpretation of Virgo's [O~III] $\lambda 5007$ 
sources as intergalactic planetary nebulae has been called into 
question by the {\sl VLT/FORS\/} spectroscopy of \citet{kud00}.
Although spectroscopic observations by \citet{free00} appear to confirm 
that most of the emission line sources identified by \citet{ciar98} and 
\citet{fcj98} are, indeed, [O~III] $\lambda 5007$ emitters in nearby 
intracluster space, the Virgo PN candidates observed by Kudritzki \tetal are 
all background galaxies.  If most of the emission-lines sources found in the
[O~III] surveys are also background galaxies, then many of the conclusions
recently reached about intracluster starlight in Virgo and Fornax are 
incorrect, and the usefulness of planetary nebula surveys in intracluster space 
has to be reconsidered.

Here we report the results of two experiments designed to estimate the
number of background galaxies expected in surveys of Virgo and Fornax
intracluster planetary nebulae (IPNe).  First, we describe spectroscopic 
observations of five planetary nebulae in the field of M87 that have been 
labeled by \cite{ciar98} as ``overluminous'' and therefore foreground 
to M87.  We show that a composite spectrum made from these sources contains
both [O~III] $\lambda 5007$ and [O~III] $\lambda 4959$ in the appropriate
ratio.  Consequently, it is likely that the majority of the overluminous 
objects in this field belong to the cluster.   We then report the results
of a 0.13 deg$^2$ on-band/off-band photometric survey at 5019~\AA\ of a 
``blank field'' set well away from any nearby galaxy or cluster.   We show
that, although the density of ``PN-like'' sources in the field is not 
negligible, it is still five times smaller than the density derived for 
[O~III] sources in Virgo.  Thus, surveys for intracluster PNe remain a 
viable method for measuring the amount of stars stripped out of galaxies.  
Finally, we discuss the nature of the contaminating objects, and, using 
arguments based on emission line equivalent widths, conclude that most of 
these sources are emission-line galaxies at $z = 3.13$.  Our results support
the conclusions of \citet{hcm} that Ly$\alpha$ galaxies with
monochromatic fluxes of $\lesssim 6 \times 10^{-17}$~ergs~cm$^{-2}$~s$^{-1}$ 
and observed equivalent widths greater than 80~\AA\  are fairly common,
although the density we derive is only one-half the Hu \tetal value.

\section{Spectroscopy of Candidate Intracluster Planetary Nebulae}

In their on-band/off-band [O~III] $\lambda 5007$ survey of planetary
nebulae in the field of M87, \citet{ciar98} found that the
planetary nebula luminosity function (PNLF) of the galaxy is distorted 
relative to the PNLFs of other galaxies.   Specifically, in the outer part
of the galaxy, M87's PNLF does not have a sharp cutoff at its bright end;
instead, the galaxy appears to contain a population of ``overluminous''
objects that are up to $\sim 0.6$~mag brighter than the PNe seen in other
galaxies.  Because the number of these overluminous PNe scales with the
survey area, and not galactic surface brightness, Ciardullo \tetal
concluded that these bright [O~III] $\lambda 5007$ sources are not 
associated with M87.  Instead, Ciardullo \tetal hypothesized that the
overluminous PNe belong to the Virgo Cluster as a whole, and 
are projected up to $\sim 3$~Mpc in front of M87.  

To test this hypothesis, we used the {\sl Hydra\/} multi-fiber spectrograph
on the Kitt Peak {\sl WIYN\/} telescope to obtain spectra of
some of the overluminous PN candidates.  Our specific spectrograph setup 
included {\sl Hydra's\/} 2-arcsec red sensitive fiber cables, the bench 
spectrograph's 285 mm focal length camera, a SITe $2048 \times 2048$~CCD 
detector (4.3~e$^-$ readnoise, gain of 1.7~e$^-$~ADU$^{-1}$), and an 
860~lines~mm$^{-1}$ grating blazed at $30\fdg 9$.  When used in second order, 
this grating-camera-detector combination yielded spectra with a dispersion of 
0.47~\AA~pixel$^{-1}$ and a resolution of 0.77~\AA\ over the wavelength range 
between 4500~\AA\ to 5500~\AA{}.  The total observation time for the setup was 
three hours; to facilitate the removal of cosmic rays, the observations were 
split into three one-hour exposures.

Nine objects with monochromatic [O~III] $\lambda 5007$ fluxes greater than
$1.0 \times 10^{-16}$~ergs~cm$^{-2}$~s$^{-1}$ ($m_{5007} < 26.25$) were
chosen for study.  Since M87's PNLF cutoff occurs at $m_{5007} = 26.35$
\citep{ciar98,jcf90}, these PN candidates are all significantly 
``overluminous'' (by $\sim 2\sigma$) with respect to the galaxy, and can 
be interpreted as foreground intracluster stars.  Unfortunately, due to a 
combination of factors, including astrometric and fiber-positioning errors, 
only five of the emission-line objects were detected.   

The multifiber spectra were extracted, flatfielded, and wavelength 
calibrated using the IRAF package {\tt dohydra}.  A flatfield image
taken at the beginning of the night served as the reference for
tracing the spectra on the CCD frame; the wavelength calibration was
performed using copper-argon comparison arcs taken before and after
the exposure sequence.  Sky subtraction was accomplished using the summed
spectra of 49 sky fibers randomly placed around the galaxy.   Since the
sky in this wavelength region is faint and our observations were taken at
moderately high ($R \sim 6000$) resolution, our results are independent of
the accuracy of this subtraction.

The overluminous PN candidates are listed in Table~1, using the identification
numbers of \citet{ciar98}. Figure~1 displays the recorded spectra in 
the wavelength region between 4940~\AA\ and 5060~\AA.  As the figure 
illustrates, the signal-to-noise of the spectra is extremely low; in all 
cases, we have only a bare detection of a single, unresolved emission line.
The fact that the emission lines are unresolved does suggest that the objects 
are Virgo PNe, rather than background galaxies (or quasars), since the 
velocity widths of the latter objects would likely be greater than our 
$\sim 50$~\kms\ resolution.  However, for any individual object, this 
possibility cannot be excluded.

Nevertheless, we can use the procedure of \citet{free00} to
{\it statistically\/} show that the detected emission line is [O~III]
$\lambda 5007$.  To do this, we start by assuming that the identification of
[O~III] $\lambda 5007$ is correct, and then shift each spectrum back into its
rest frame.  We then scale the data so that all the $\lambda 5007$ lines have 
equal weight, and co-add the spectra to create a single composite spectrum.
This spectrum is displayed in Figure~2.  As the figure illustrates,
the increased signal-to-noise brings out an additional feature: [O~III] 
$\lambda 4959$.  If all five overluminous objects are indeed, intergalactic
PNe, then this companion line should be $1/3$ the strength of $\lambda 5007$; 
if some of the objects are background galaxies, then the line ratio should be 
reduced by the fraction of contaminants.  

The ratio of $\lambda 4959$ to $\lambda 5007$ in Figure~2 is 
$0.31 \pm 0.03$, where the quoted error represents our estimate of the
measurement uncertainty, and neglects factors associated with the
changing efficiency of the spectrograph.  Since this number is extremely
close to $1/3$, it is likely that all of the overluminous $\lambda 5007$
sources are planetary nebulae: at most, only one of the five objects
can be a contaminant.  Moreover, since the sources are too bright to be PNe 
associated with M87, the best hypothesis remains that these objects belong to 
the Virgo Cluster as a whole, and are located on the near side of the cluster.

\section{[O~III] $\lambda 5007$ Imaging of a Blank Field}

An alternative method of determining the importance of contamination in
IPNe surveys is through the analysis of a control region.  By measuring
the surface density of point-like emission-line sources in the field, we can
directly estimate the amount of contamination expected in cluster surveys.  
To perform this ``blank field'' experiment, we chose a high-latitude 
($b = -49^\circ$) location in the southern hemisphere, 
$\alpha(2000) = 4^{\rm h} 01^{\rm m}$, $\delta(2000) = -39^\circ 42\arcmin$.   
This region of sky is located well away from any nearby cluster: it is 
$16^\circ$ from the center of Fornax, and $2\fdg 4$ from the nearest (small) 
galaxy in the Tully Catalog \citep{tul88}.  Aside from one faint radio source 
\citep[PMN J0401$-$3944;][]{gw93} and one $z = 0.26$ quasar 
\citep[Osmer 14;][]{o82} no remarkable objects are known to be located within 
the $30\arcmin \times 30\arcmin$ field of our camera.  

According to the H~I and galaxy count data of \citet{bh82}, the foreground
galactic reddening in the region is $E(B-V) = 0.0$; from the DIRBE/IRAS
data of \citet{sfd98}, this number is $E(B-V) = 0.029$. For simplicity, in 
this paper we assume that the extinction is identically zero.

We observed the blank field on the nights of 1998 Nov 20, 21, 22, and 23
with the Big Throughput Camera (BTC) on the Cerro Tololo 4-m telescope.  This
mosaic CCD system consists of a square array of four $2048 \times 2048$ 
Tektronix CCDs, each of which has a readnoise of $\sim 5$~e$^-$, a gain of 
2.4~e$^{-}$~ADU$^{-1}$ and a scale of $0\farcs 43$ per pixel.  The total area 
included on all four chips is 0.23 deg$^2$, but due to the dithering 
procedure used to improve the flatfielding, the actual area surveyed was only 
0.215 deg$^2$.  When we exclude the data of CCD~\#2, which has a slightly
lower (but highly variable) quantum efficiency, and eliminate the areas around
saturated stars and bad pixels, our effective survey region is further reduced 
to 0.13 deg$^2$.  Still, this is over six times larger than that of
the deep, narrow-band Ly$\alpha$ surveys of \citet{hcm} and \citet{steid00}.

The on-band filter used for our experiment was the [O~III] \#2 Mosaic
filter of Kitt Peak; this filter has a central wavelength of 5019~\AA\
and a full-width-at-half-maximum (FWHM) of 55~\AA\ at ambient temperature
in the converging beam of the telescope.  The off-band filter was Kitt Peak
Mosaic filter [O~III]+29, whose central wavelength and FWHM are 5305~\AA\
and 241~\AA, respectively.  The total on-band exposure time for our four 
nights of observations was 285 minutes; this was split into four 1-hr exposures
plus one 45-min exposure. Our off-band data consisted of 5 15-min exposures,
for a total integration time of 75 min.  All these data were taken under
photometric conditions in $1\farcs 1$ seeing.  

The procedures used to reduce the data, identify emission-line objects,
and measure their brightnesses, were identical to those used to search for
intracluster planetary nebulae, and are fully described in \citet{feld00}.
After de-biasing and flatfielding the data, the on-band and off-band frames
were co-added to create master images that were clipped of cosmic-rays.
The frames were then searched for PN-like candidates in two different
ways using the DAOPHOT and PHOT packages within IRAF{}.  The first method
involved performing photometry of all point sources on the frames, 
and constructing a color-magnitude diagram from the results.  Objects
with highly negative on-band minus off-band colors were flagged as 
PN-like sources (cf.~Figure~3).   In the second method, candidate sources
were identified by searching for positive residuals on a ``difference image''
made by subtracting a scaled version of the off-band image from the on-band 
frame.  These two techniques complement each other, in that each detects
objects that the other method does not.  Because the color-magnitude algorithm 
is susceptible to confusion by continuum sources near target objects (blends), 
the former analysis misses $\sim 5\%$ of the emission-line sources, even at
bright magnitudes.  Conversely, the subtraction technique detects virtually
all the bright emission-line objects, but loses some of the fainter sources
due to the larger noise of the difference frame.  As we describe below,
these two techniques, coupled with our chosen detection threshold for DAOFIND, 
optimized our our ability to detect real emission-line sources, and kept the 
number of false detections at a manageable level.

After analyzing our on-band/off-band frames with both methods, a series
of routines were run to purge the candidate list of false detections.  First,
we removed objects which fell within the excluded regions around saturated
stars and near bad pixels.  We then ignored all objects for which the final
final detected signal-to-noise was less than four; such objects are almost
always random noise fluctuations \citep[cf.][]{ciar87,hui93}.  Finally, we
removed probable cosmic rays from our list by going back to the individual
exposures which created the master on-band frames.  Using apertures of the 
order of the seeing FWHM, we measured the magnitude of each candidate object 
on each individual frame.  If, after correcting for sky transparency, one of 
the measurements was $5 \sigma$ higher than the mean determined from the 
magnitudes on the other four frames, the object was deleted from the analysis. 

Once the false detections and cosmic rays were eliminated, the 
remaining PN-like candidates were screened by eye to create the final list 
of $\lambda 5019$ sources.  To be included in the list, an object had to be 
spatially unresolved on the on-band image, and completely invisible on the 
off-band frame.  A total of nine such ``PN-like'' objects were found by our 
procedure.

To confirm the robustness of our algorithms, we tested our results in three
different ways.  First, we added artificial stars to the on-band (but not the
off-band) image using the ADDSTAR task within DAOPHOT{}.  We then ran our 
detection algorithms, and computed the fraction of objects recovered as a 
function of magnitude.  Since our algorithms can work no better than that of 
the star-finding algorithm \citep[DAOFIND;][]{stet87}, this analysis places a 
limit on the effectiveness of the search.  The result of this simulation for 
one of the BTC CCDs is shown in Figure~4.  As the figure illustrates, we are 
90\% complete down to an instrumental magnitude of $m_{inst} = 27.74$, and 
50\% complete to $m_{inst} = 28.19$.   For consistency with other planetary 
nebula surveys, we adopt the 90\% completeness limit, which is equivalent
to a signal-to-noise of nine \citep{ciar87,hui93}, as this survey's limiting 
magnitude.

As a second test, we ran our automated detection algorithms on frames 
of the Virgo Cluster that had been visually blinked as part of previous 
surveys for intracluster planetaries \citep{fcj98, fcj01}.
Our new algorithms recovered all of the PN candidates above the 
DAOFIND completeness limit and 92\% of all the previously known sources.
Finally, two of us (J.J.F. and K.K.) independently blinked one of the BTC CCD
fields by eye, and compared our results to those of the automated 
algorithm.  In both cases, all objects brighter than $m_{inst} = 27.7$ that 
were found by eye were also recovered by the computer.  We are therefore 
confident that our detection scheme is the best possible for finding
faint emission-line sources in a blank field.

Once the objects were found, their equatorial positions were derived with
respect to the reference stars of the USNO-A 2.0 astrometric catalog
\citep{mon98,mon96}.  The measured residual of our plate solution
was $\sim 0\farcs 2$, a number slightly less than the $0\farcs 25$
external error associated with the catalog.  On-band and off-band photometry
was accomplished relative to bright field stars via the aperture 
photometry routines within IRAF\null.  These data were then placed
on the standard AB system by comparing large aperture measurements 
of field stars with similar measurements of the \citet{stone77} and 
\citet{sb83}  spectrophotometric standards Feige~110, LTT~1020, LTT~3218, 
EG~21, and LTT~2415.  The dispersion in the photometric zero point computed 
from these stars was $0.03$~mag.  Finally, we converted the on-band AB 
magnitudes to $\lambda 5007$ monochromatic fluxes using the techniques outlined
in \citet{jqa87}, under the assumption that the wavelength of the detected 
emission line lies near the bandpass center of the interference
filter.  For consistency with papers on extragalactic planetary nebulae,
we define
\begin{equation}
m_{5007} = -2.5 \log F_{5007} - 13.74
\end{equation}
where $F_{5007}$ represents monochromatic flux in ergs~cm$^{-2}$~s$^{-1}$.

When placed on this standard system, the instrumental completeness limit of
$m_{inst} = 27.7$ corresponds to an AB magnitude of $m_{AB}^{\rm on} = 24.3$,
or a $\lambda 5007$ magnitude of $m_{5007} = 27.0$.  A similar analysis for the 
continuum frame reveals that the limiting AB magnitude of our off-band data is
$m_{AB}^{\rm off} = 24.7$.  Coupled together, these two numbers imply that our 
$\lambda 5007$ survey should be complete to a monochromatic flux of 
$5.0 \times 10^{-17}$~ergs~cm$^{-2}$~s$^{-1}$, which corresponds to an 
equivalent limit of $W_{\lambda} > 82$~\AA\ in the observer's frame.

\section{The Density of Background Sources}

Those objects which, had they been located in a cluster environment, 
would have been mistaken for intracluster planetary nebulae, are listed 
in Table~2.  All the detected point-sources have {\it
extremely large\/} equivalent widths: the smallest value is 212~\AA,
and the median equivalent width is 273~\AA.  These numbers are far above
our detection threshold of 82~\AA, and demonstrate that the number of
PN-like sources in the blank field is not sensitive to this quantity.  The 
most important feature of the table, however, is its rather small number
of objects.  Only nine point sources brighter than our limiting
magnitude of $m_{5007} = 27.0$ were detected in the 0.13 deg$^2$
region.  This compares to nine objects brighter than $m_{5007} = 26.8$
in the lowest density Virgo field observed by \citet{fcj98}.
Considering that the Feldmeier \tetal field is less than
half the area of this survey, the data imply that surveys for intracluster
PNe are a viable way to probe the distribution of intracluster
starlight.

The results, however, do suggest that some of the recent conclusions about
intracluster stars in Virgo and Fornax may be incorrect.  For example,
\citet{fcj98} used the apparent luminosity of Virgo's 
brightest [O~III] $\lambda 5007$ sources and the planetary nebula luminosity 
function, to place an upper limit on the distance to the front edge of the 
system.  Their distance of $\sim 11.8$~Mpc implied that Virgo is extremely 
elongated along the line-of-sight.  Based on our blank field study, it is clear
that field objects with large equivalent widths and line fluxes as large
as $\sim 7 \times 10^{-17}$~ergs~cm$^{-2}$~s$^{-1}$ do exist, and can 
be mistaken for intracluster planetaries.  Thus, structural analyses that 
depend on one or two extremely bright objects may be incorrect.   In the
case of the M87 field, the spectroscopy of Section~2 demonstrates that 
the intracluster environment is indeed elongated:  since the absolute
magnitude of the [O~III] $\lambda 5007$ PNLF bright-end cutoff is $M^* = -4.5$
\citep{ciar89}, the apparent magnitudes of Table~1 imply that the 
front side of Virgo is no more distant than 12.8~Mpc \citep[\ie $\sim 1.6$~Mpc
in front of M87;][]{ciar98}.  Unfortunately, without spectroscopy, 
similar limits cannot be placed on other regions of the cluster.

To quantify the fraction of contamination expected, Table~3 compares the
density of intracluster [O~III] sources detected in Fields 2-6 of
\citet{fcj98,fcj01} to the results of this blank field experiment.  
The limiting magnitudes of the Feldmeier \tetal fields are generally
not as deep as the data presented here ($m \lesssim 26.8$), so the Virgo 
results have been extrapolated to a depth of $m_{5007}= 27.0$ using the 
standard form of the PNLF:
\begin{equation}
N(M) \propto e^{0.307M} \, [1 - e^{3(M^{*}-M)}]
\end{equation}
Although the contamination rate fluctuates from field to field, Table~3
demonstrates that $\sim 20\%$ of the emission-line objects found by Feldmeier 
\tetal are likely to be interlopers.  Unfortunately, it is difficult to
define this fraction more precisely.  Due to the small number of objects 
detected in the blank field survey, the derived contamination rate is 
uncertain by at least 33\%.  Moreover, if the contaminating sources are,
indeed, extragalactic background objects, some field-to-field fluctuation
might be expected to arise from large-scale structure.  Nevertheless, our
photometrically derived contamination fraction is in good agreement with 
the value of 26\% determined via spectroscopy of individual
IPN candidates \citep{free00}.

A $\sim 20$\% contamination rate, coupled with the distribution of magnitudes
given in Table~2, explains why \citet{kud00} found a large fraction of 
contaminants in their spectroscopic survey, but \citet{free00} did not.  From
the table, it is apparent that there are more faint emission-line sources 
in our blank sky field than  bright sources.  A maximum-likelihood analysis 
of the nine data points suggests that, in the magnitude range $m_{5007} < 27$, 
a power-law $\log N(m) = a + b m$, with slope $b \approx 1^{+0.5}_{-0.3}$ 
adequately represents the differential source counts.  If this behavior
continues to fainter magnitudes, then the apparent discrepancy between
the results of Kudritzki \tetal and Freeman \tetal is simply a function
of which objects the authors chose to observe.  Of the nine objects 
observed by Kudritzki {\refit et al.,} two are extremely bright,
with [O~III] $\lambda 5007$ magnitudes more than 0.7~mag brighter than the
nominal PNLF cutoff.  Since the PNLF's decline at bright magnitudes is
extremely rapid, blank field sources dominate at these magnitudes.  Moreover, 
four of the Kudritzki \tetal targets are extremely faint, with $m_{5007} > 
27.5$.   In this regime, the slope of the PNLF has reached its limiting value 
of 0.133.  Since this number is $\sim 7$ times shallower than the background
source count slope, a significant amount of contamination is again expected.  
Between these two extremes, however, the density of background objects is much 
less than the density of IPNe, and it is this range that Freeman \tetal 
performed their observations.  

If we adopt a slope of $b \sim 1$ for the luminosity function of the 
background objects, and normalize the counts so that the background 
contributes $\sim 20\%$ of the $m_{5007} < 27$ sources in Virgo's 
intracluster space, then by $m_{5007} \sim 27.6$, the density of interlopers
is more than half that of the intracluster stars.  Of course, this
result depends sensitively on the extrapolated slope of the background
source counts.  Nevertheless, because the faint-end slope of 
the PNLF is shallow, we can expect that at some point past $m_{5007} = 27$, 
the background sources will overwhelm the planetaries.

Our estimate of the contamination rate moves PN-based measurements 
of Virgo's intracluster stars into better agreement with that determined
from {\sl HST\/} red giant star counts \citep{ftv98}.  However, because
the vast majority of PN candidates brighter than $m_{5007} = 27.0$ are still 
likely to be genuine, a factor of $\sim 2$ discrepancy between the two
estimates remains.  One possible explanation for the difference may 
be small-scale structure in the distribution of the intracluster stars 
themselves.  Observationally, there is evidence that the IPN are scattered 
non-uniformly in the Virgo cluster \citep{feld00} and most theories for 
the creation of intracluster stars predict that the distribution of such 
objects should have filamentary structure \citep{moore96}.  Since the 
{\sl HST\/} field is much smaller ($\sim 5$~sq.~arcmin) than the 
IPN fields ($\sim 200$~sq.~arcmin), it is entirely possible that the
difference between the IPN and {\sl HST\/} estimates for Virgo's intracluster
star population reflect real fluctuations in the stellar background.  

\section{What are the Contaminating Sources?}

To be detected via our on-band/off-band interference technique, an object must
have an extremely strong emission line, with an observed equivalent width of
$W_\lambda > 82$~\AA.  If the line is not [O~III] $\lambda 5007$, then the
two most plausible alternatives are [O~II] $\lambda 3727$ at $z = 0.35$ (rest
frame equivalent width $W_\lambda > 60$~\AA) or Ly$\alpha$ at $z=3.13$
($W_\lambda > 20$~\AA).  The former option is possible, though unlikely.  Most
of our emission-line sources have putative [O~II] rest frame equivalent widths
greater than 150~\AA; the redshift surveys of \citet{cow96}, \citet{ham97},
and \citet{hog98} have yet to find such extremely strong [O~II] emitters
\citep[but see][]{stern00}.  Moreover, if the emission-line sources are really
[O~II] galaxies at $z = 0.35$, then these objects must have absolute magnitudes
between $-15.3 < M_B + 5 \log h < -15.0$.  (The bright limit comes from the
limiting magnitude of our off-band frame, while the faint limit is derived 
from our on-band flux limit of $5 \times 10^{-17}$~ergs~cm$^{-2}$~s$^{-1}$
and the observed $\sim 150$~\AA\ upper limit for [O~II] equivalent widths.)  
If we integrate the luminosity function for low-redshift [O~II] emission-line 
galaxies \citep{ellis96, ssh97} between these two limits, and normalize to the
volume of space surveyed by our 55~\AA\ FWHM filter ($\sim 800~h^{-3}$~Mpc$^3$ 
at $z = 0.35$), then we find that $\sim 4$~[O~II] emission line galaxies should
lie within our survey region.   Since only $\sim 10\%$ of faint ($M_B > -19$) 
[O~II] emission line galaxies have equivalent widths greater than our 60~\AA\ 
rest frame survey limit \citep{cow96, ham97, hog98}, the implication is that
$\lesssim 1$ of our strong emission-line sources is an [O~II] galaxy.

On the other hand, it is likely that at least some of our emission-line
sources are Ly$\alpha$ objects.   In their spectroscopic follow-up of 
candidate intracluster planetary nebulae in Virgo, \citet{kud00}
found nine such objects.  Similarly, in their deep narrow-band and 
spectroscopic searches of a 75~square arcmin region, \citet{hcm}
found that Ly$\alpha$ sources with observed equivalent widths
$W_\lambda > 90$~\AA\ are not rare.  Specifically, these authors found that,
down to a limiting flux of $\sim 1.5 \times 10^{-17}$~ergs~cm$^{-2}$~s$^{-1}$, 
the density of Ly$\alpha$ emitters at $z \sim 3.4$ is $\sim 15,000$~deg$^{-2}$ 
per unit redshift interval.  Since only five out of the twelve emitters 
tabulated by \citet{ch98} have fluxes greater than our detection limit 
of $5 \times 10^{-17}$~ergs~cm$^{-2}$~s$^{-1}$, their observations suggest 
that we should be able to detect $\sim 38$ such objects in our survey region.

In order to compare our results to the Cowie \tetal numbers, we expanded our
search criteria to include both resolved sources and sources detected in
continuum.  To do this, we followed the prescription of \citet{herr01}
and modified DAOFIND's FWHM convolution kernel so that our program could detect
non-stellar objects.  We then re-ran our search algorithms on the BTC images, 
and flagged all objects (both stellar and non-stellar) with fluxes above the 
completeness limit and with equivalent widths greater than 82~\AA.  (For 
objects with no continuum detection, we used the off-band frame's $1 \, \sigma$
upper limit in our calculation of equivalent width.)  The 12 additional objects
found in this search are listed in Table~4, and plotted as open circles in
Figure~3.  Figure~5 displays the on-band, off-band, and difference images of 
a sample of our strong line-emitters (both PN-like and otherwise), and 
finding charts for all the sources listed in Tables~2 and 4 are given in 
Figure~6.  In total, we detected 21 strong emission-line sources in our 0.13 
deg$^2$ survey region.  Of the 12 additional sources, five are well-resolved
with FWHM sizes in the range between $\sim 1\farcs 1$ and $\sim 2\farcs 2$, 
and another three are marginally resolved.  This size distribution (13 point
sources, 3 marginally resolved objects, and 5 clearly resolved galaxies) 
appears roughly consistent with the distribution of half-light radii observed 
for Lyman-break galaxies \citep[cf.][]{low97}.  However, higher resolution
images are needed to confirm this result.

Our surface density, coupled with the $\Delta z = 0.046$ bandpass of our 
filter, implies that the surface density of Ly$\alpha$ sources with
monochromatic fluxes greater than $5 \times 10^{-17}$~ergs~cm$^{-2}$~s$^{-1}$ 
is $\sim 3500$~deg$^{-2}$~per unit redshift interval.  This value is 
marginally lower (by $1.4 \, \sigma$) than the value derived by \citet{hcm}. 
The discrepancy certainly is not serious, especially since both surveys may 
be affected by large-scale structure.  By counting Lyman-break galaxies in 
$9\arcmin \times 9\arcmin$ regions, \citet{adel98} has estimated the 
cell-to-cell variance of $z \sim 3$ Lyman-break galaxies to be 
$\sigma^2 \sim 1.3 \pm 0.4$.  However, the large-scale distribution of 
Ly$\alpha$ emitters may be quite different, and significant excursions 
from the mean are possible.  For example, in their 5-m survey of a 
78~square arcmin region of the sky, \citet{steid00} detected 
$\sim 25$ $z=3.09$ emission-line sources above our detection threshold.   
When scaled to the bandwidth of our $\lambda 5007$ filter, their data give a 
Ly$\alpha$ galaxy surface density of $\sim 17,000$~deg$^{-2}$ per unit 
redshift interval, a number almost five times larger than observed here.  
However, the Steidel \tetal survey was directed at a large structure of 
galaxies, whose volume density is $\sim 6.0 \pm 1.2$ times higher than 
average.  Thus, their result is consistent with our observations.


For an $H_0 = 65$~km~s$^{-1}$~Mpc$^{-1}$, $q_0 = 0.5$ universe, our data
imply that the space density of $z=3.13$ galaxies with Ly$\alpha$(EW) 
$> 21$~\AA\ and $L({\rm Ly}\alpha) > 2.2 \times 10^{42}$~ergs~s$^{-1}$ 
is $9.3 \times 10^{-4}$~galaxies~Mpc$^{-3}$.  Note that these objects are
undergoing a significant amount of star formation.   If we neglect internal 
extinction and assume \citep[as do][]{hcm} that the Ly$\alpha$ to H$\alpha$ 
ratio of these objects is 8.7 \citep{b71}, then $10^{42}$~ergs~s$^{-1}$
in Ly$\alpha$ corresponds to a star formation rate of $1 M_\odot$~yr$^{-1}$
\citep[cf.][]{ken83}.  This implies that all but one of the objects detected
in this survey have star formation rates between 1 and $8~M_\odot$~yr$^{-1}$.
This is consistent with the $z > 3$ star formation rates derived
from the rest-frame UV flux measurements of \citet{steid98}.

Figure~7 plots the emission-line luminosity function of strong emission-line
sources, under the assumption that the objects are, indeed, Ly$\alpha$
galaxies at $z=3.13$.  If we integrate the function over the range displayed 
in the figure, we obtain a measured star formation rate density of 
$\sim 0.003~M_\odot$~yr$^{-1}$~Mpc$^{-3}$.  If we then extrapolate
the luminosity function to the approximate limiting magnitude of Hu \tetal 
using the assumption that the luminosity function flattens past
$L \sim 2 \times 10^{42}$~ergs~s$^{-1}$ in a manner similar to a \citet{sch76}
function, then the total star formation rate density of
emission line galaxies brighter than $\sim 7 \times 10^{41}$~ergs~s$^{-1}$
is $\sim 0.004~M_\odot$~yr$^{-1}$~Mpc$^{-3}$.  This value is roughly
one-tenth that derived from Lyman-break galaxies \citep{steid99}.

Our value for the space density of Ly$\alpha$ emitters and their implied
star formation rate density is a factor of $\sim 2$ smaller than that
derived by \citet{hcm}.  Since the Hu \tetal data are for a slice of the 
universe at $z=3.4$, it is possible to interpret this decrease as being
due to evolution.  However, given the small difference in time 
between these two epochs ($\sim 0.2$~Gyr) and the small number of galaxies 
detected, it is far too early for such a conclusion.


\section{Conclusions}

By studying a blank field located well away from any galaxy or cluster, we have
found that surveys for intracluster planetary nebulae are not pristine.  At
monochromatic fluxes fainter than $\sim 6 \times 10^{-17}$~ergs$^{-2}$~s$^{-1}$
there is a non-negligible amount of contamination from field objects.
The most likely source of this contamination is Ly$\alpha$ galaxies at 
$z \sim 3.13$; the data therefore suggest that the same techniques that
allow 4-m telescopes to map out the distribution of intracluster PNe,
can be used to determine the distribution of starbursting galaxies at 
high redshift.

Nevertheless, our observations confirm the presence of intracluster PNe
in Virgo.  The amount of background contamination is not nearly enough
to explain the number of emission-line sources found in wide-field
surveys of Virgo; based on our blank-field data, the contamination fraction
is only $\sim 20\%$.  This low number is confirmed by our spectroscopy of
candidate intracluster PNe in the field of M87:  the presence of 
[O~III] $\lambda 4959$ in a composite spectrum of five ``overluminous'' 
objects implies that at least four of the objects are true planetary
nebulae.   Surveys for intracluster planetaries at magnitudes brighter
than $m_{5007} \sim 27$ therefore remain an effective way of probing the 
dynamical history of nearby clusters.

\acknowledgements

We thank K. Freeman, R. Kudritzki, and R. M\'endez for useful discussions
about the results of their spectroscopy of intracluster PN candidates,
and J.A. Tyson for his assistance with the BTC camera.
This research made use of NED, the NASA Extragalactic Database, and was 
supported in part by NSF grants AST 92-57833 and AST-9529270.

\pagebreak

\pagebreak
\figcaption[]
{The spectra of five ``overluminous'' emission line sources in the field 
of M87; the ID numbers come from \citet{ciar98}.  Although the spectra
have very low signal-to-noise, a single unresolved emission line, which is 
presumed to be [O~III] $\lambda 5007$ is clearly visible.  The fact that the
emission line is unresolved at $\sim 50$~\kms\  resolution suggests that the 
line is not [O~II] $\lambda 3727$ or Ly$\alpha$ emission from a background
galaxy.}

\figcaption[]
{A composite spectrum of the five ``overluminous'' emission line sources,
made using the assumption that the observed emission line is, indeed,
[O~III] $\lambda 5007$.  In the spectrum [O~III] $\lambda 4959$ is 
clearly visible, with a strength that is approximately 1/3 that of
[O~III] $\lambda 5007$.  The second line confirms that the objects are
local to the Virgo Cluster, and not background galaxies.}

\figcaption[]
{Excess emission in the narrow-band $\lambda 5019$ filter over the continuum
for objects in our BTC field.  The abscissa gives the monochromatic
$m_{5007}$ magnitude, while the ordinate shows the difference between the
sources' on-band and off-band AB magnitudes.  Our on-band completeness
limit of $m_{5007} = 27.0$ is represented by a vertical line; our
equivalent width limit of 82~\AA\ is shown via the horizontal line.  The
dotted line represents an observer's frame equivalent width of 300~\AA,
and is shown for reference.  The curve illustrates the expected $1 \, \sigma$ 
errors in the photometry.  The emission line sources discussed in the text 
are displayed as filled circles (for point-like objects) and open circles 
(for sources resolved in $1\farcs 1$ seeing).}

\figcaption[]
{The fraction of artificial stars recovered on our on-band CCD frames as a 
function of instrumental magnitude.  The open circles represent the efficiency
of the DAOFIND detection algorithm, the solid squares plot the fraction of 
recoveries from the color-magnitude method, and the open triangles show
the efficiency of detections on the difference image.   The curves are
spline fits through the data and are for illustration only.  The
solid vertical line denotes the 90\% completeness limit of $m_{inst} = 27.44$, 
while the dashed line shows the 50\% completeness limit at $m_{inst} = 28.19$.
The simulations show that the color-magnitude method is the most efficient way 
of detecting faint emission line sources, but it fails to recover $\sim 5\%$
of objects at the bright end.  Conversely, the difference method detects
all of the brighter objects, but is not as effective at finding the
faintest sources.}

\figcaption[]
{On-band, off-band, and difference images for objects 18 (top), 1 (middle)
and 15 (bottom).  Each frame is $30\arcsec$ on a side.  North is up and east 
is to the left.}

\figcaption[]
{Finding charts for the strong emission-line sources found in this
survey.  Each frame is $1\arcmin$ on a side.  North is up and east is to 
the left.}

\figcaption[]
{The luminosity function of all emission-line sources with
equivalent widths greater than 82~\AA\ in the observer's frame
(or greater than 20~\AA\ in the rest frame of Ly$\alpha$).  The error
bars in the $y$-direction represent the uncertainty due to counting
statistics; the errors in the $x$-direction reflect the bin size
(0.025 in log luminosity).  For consistency with Cowie \& Hu (1998), the
data assume an $H_0 = 65$~km~s$^{-1}$~Mpc$^{-1}$, $q_0 = 0.5$, $\Lambda = 0$
cosmology.  If we use the assumption that $10^{42}$~ergs~s$^{-1}$ of 
Ly$\alpha$ photons is produced by $1 M_\odot$~yr$^{-1}$ of star formation, 
then the integrated star formation rate density implied by the diagram is
$\sim 0.004 M_\odot$~yr$^{-1}$~Mpc$^{-3}$.
}

\clearpage
\begin{figure}
\figurenum{1}
\plotone{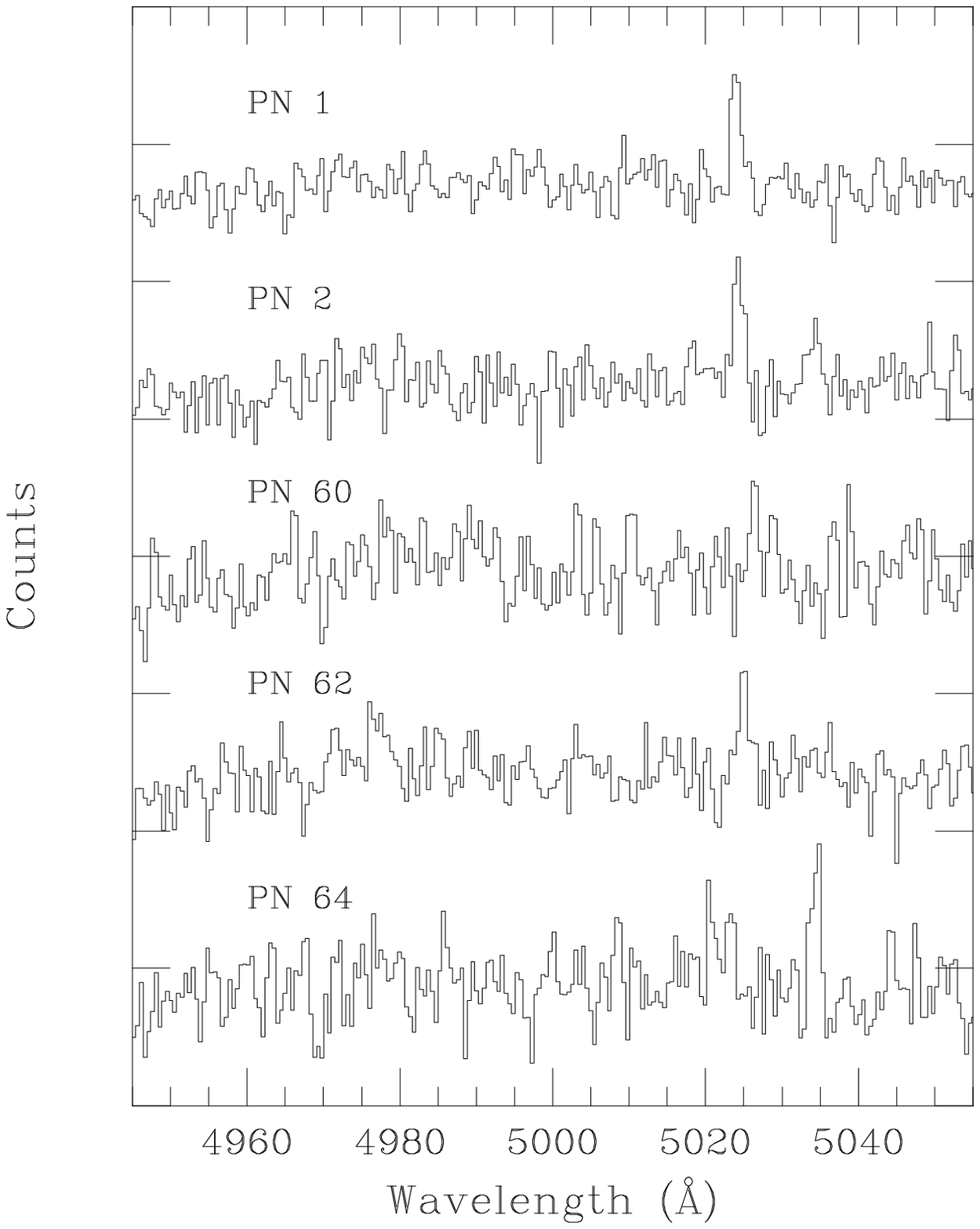}
\end{figure}
\pagebreak

\clearpage
\begin{figure}
\figurenum{2}
\plotone{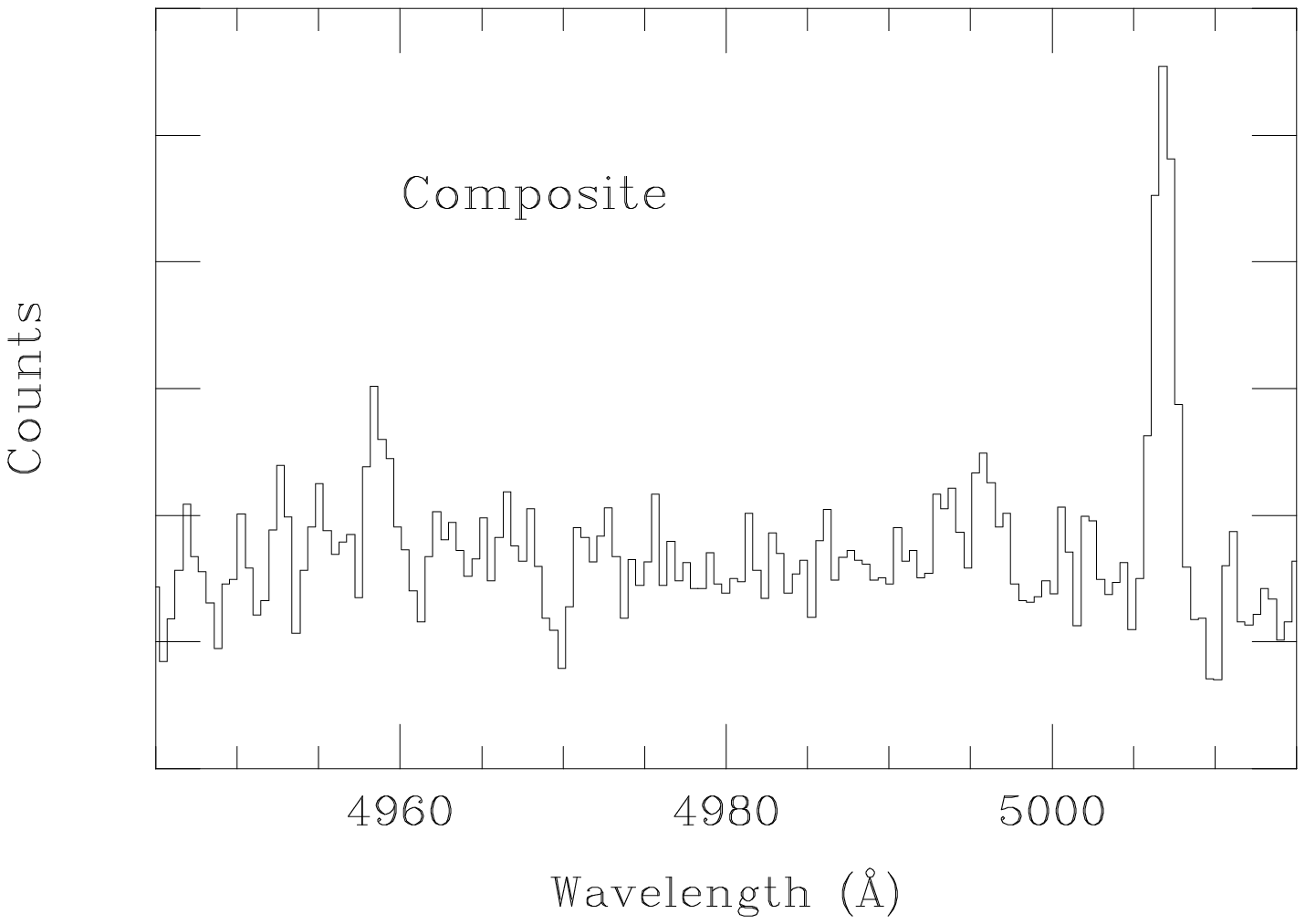}
\end{figure}
\pagebreak

\clearpage
\begin{figure}
\figurenum{3}
\plotone{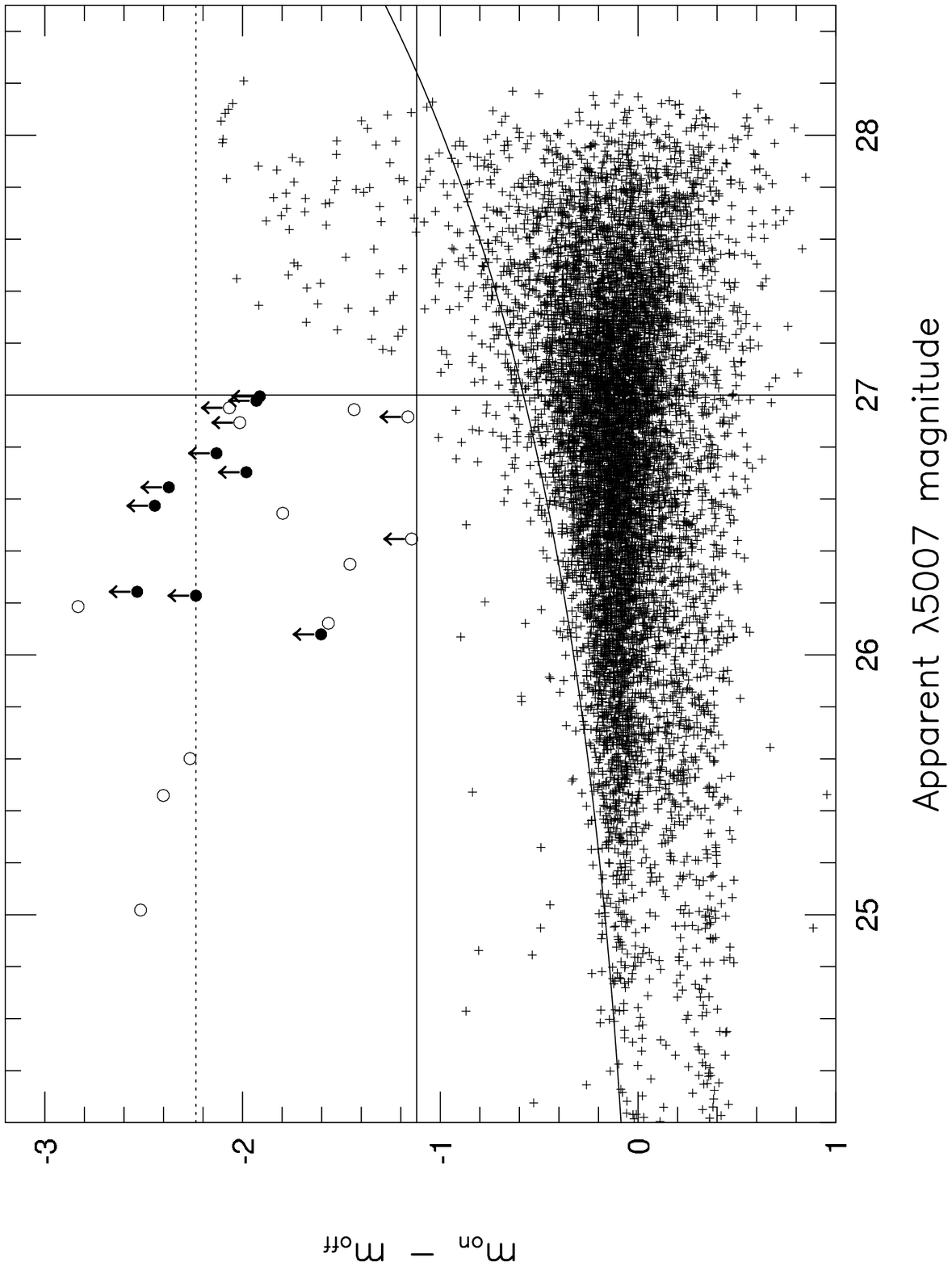}
\end{figure}
\pagebreak

\clearpage
\begin{figure}
\figurenum{4}
\plotone{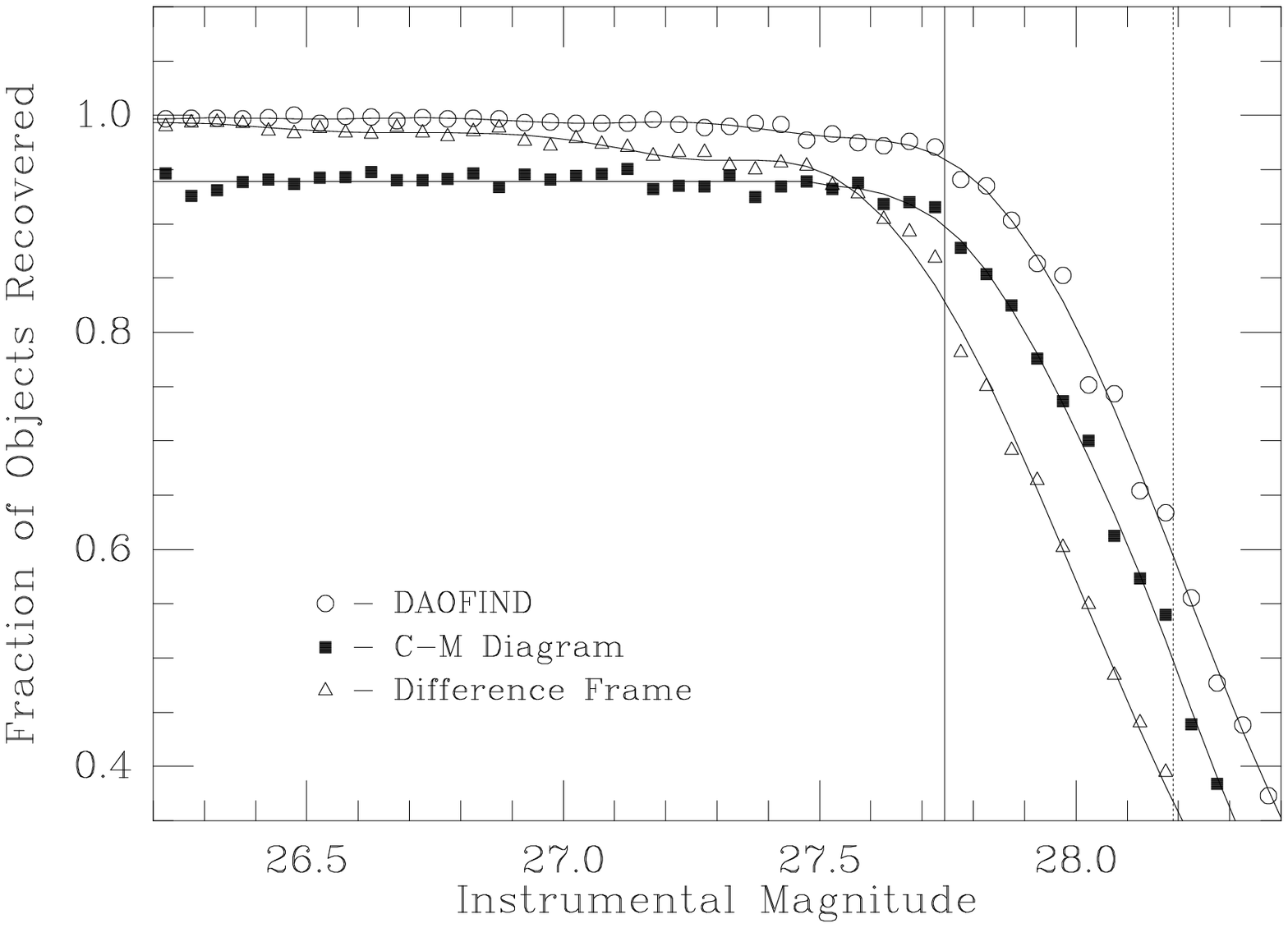}
\end{figure}
\pagebreak

\clearpage
\begin{figure}
\figurenum{5}
\end{figure}
\pagebreak

\clearpage
\begin{figure}
\figurenum{6a}
\end{figure}
\pagebreak

\clearpage
\begin{figure}
\figurenum{6b}
\end{figure}
\pagebreak

\clearpage
\begin{figure}
\figurenum{7}
\plotone{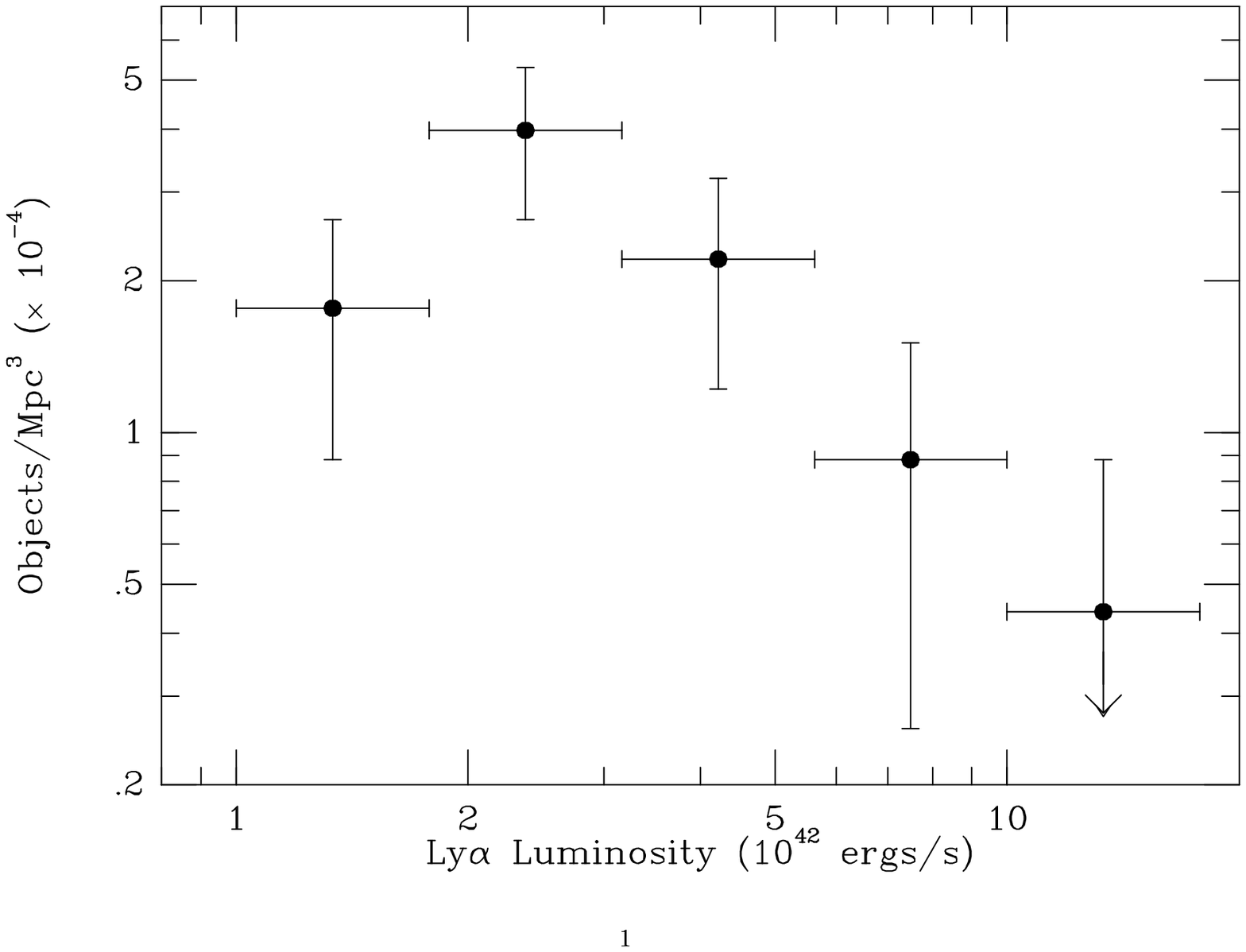}
\end{figure}
\pagebreak

\clearpage
\begin{deluxetable}{lcccc}
\tablenum{1}
\tablewidth{0pt}
\tablecaption{Velocities of M87 Overluminous PN}
\tablehead{
\colhead{ID} & \colhead{$\alpha (2000)$} & \colhead{$\delta (2000)$}
& \colhead{$m_{5007}$} &\colhead{$V$ (\kms)} }
\startdata
1   &12 30 49.38 &12 20 58.4 &25.63 &1023 \\
2   &12 30 52.44 &12 21 13.5 &26.11 &1041 \\
60  &12 30 38.95 &12 28 13.8 &26.01 &1167 \\
62  &12 30 47.36 &12 18 40.6 &26.12 &1084 \\
64  &12 30 42.14 &12 29 13.9 &26.21 &1646 \\
\enddata
\end{deluxetable}

\begin{deluxetable}{cccccc}
\tablenum{2}
\tablewidth{0pt}
\tablecaption{Blank Field ``PN-like'' Emission-Line Objects}
\tablehead{
&&&&$\lambda 5007$ Flux &Observed \\
\colhead{ID} &\colhead{$\alpha (2000)$} &\colhead{$\delta (2000)$}
&\colhead{$m_{5007}$} &\colhead{($10^{-17}$~ergs~cm$^{-2}$~s$^{-1}$)}
&\colhead{\phantom{$>$}EW (\AA)} }
\startdata
 1 &4 01 33.78 &$-39$ 36 03.7 &26.23 &10.3 &$> 300 \pm 81$ \\
 2 &4 01 19.63 &$-39$ 40 04.6 &26.70 &6.7  &$> 228 \pm 66$ \\
 3 &4 01 33.17 &$-39$ 40 10.5 &26.78 &6.2  &$> 268 \pm 83$ \\
 4 &4 00 31.83 &$-39$ 49 01.4 &26.98 &5.2  &$> 216 \pm 70$ \\
 5 &4 00 51.52 &$-39$ 44 24.0 &26.99 &5.1  &$> 212 \pm 69$ \\
 6 &4 02 54.95 &$-40$ 01 35.2 &26.08 &11.8 &$> 273 \pm 69$ \\
 7 &4 03 08.25 &$-39$ 56 43.1 &26.24 &10.2 &$> 314 \pm 83$ \\
 8 &4 00 26.29 &$-39$ 56 10.2 &26.57 &7.5  &$> 373 \pm 107$ \\
 9 &4 00 44.11 &$-40$ 08 12.3 &26.65 &7.0  &$> 346 \pm 100$ \\
\enddata
\end{deluxetable}

\begin{deluxetable}{cccccc}
\tablenum{3}
\tablewidth{0pt}
\tablecaption{Estimates of Virgo Intracluster PN Contamination}
\tablehead{
\colhead{Field} &\colhead{Area of Field} &\colhead{Limiting} 
&\colhead{Expected \# of} &\colhead{Extrapolated \# of}
&\colhead{Contamination} \\
\colhead{} &\colhead{(arcmin$^{2}$)} &\colhead{Magnitude}
&\colhead{Contaminants} &\colhead{IPN Candidates} &\colhead{Rate} }
\startdata
2       & 252 &26.8 & 4.8 & 18 & 26.7\% \\
3       & 241 &27.0 & 4.6 & 21 & 21.9\% \\
4       & 158 &26.6 & 3.0 & 21 & 14.3\% \\
5       & 200 &26.1 & 3.8 & 30 & 12.7\% \\
6       & 204 &26.8 & 3.9 & 14 & 27.3\% \\
Average &     &     &     &    & $20.6 \pm 6.9\%$ \\
\enddata
\end{deluxetable}

\begin{deluxetable}{ccccccc}
\tablenum{4}
\tablewidth{0pt}
\tablecaption{Other Blank Field Emission-Line Sources}
\tablehead{
&&&&&$\lambda 5007$ Flux &Observed \\
\colhead{ID} &\colhead{$\alpha (2000)$} &\colhead{$\delta (2000)$}
&\colhead{$m_{5007}$} &\colhead{$m_{AB}^{\rm off}$} 
&\colhead{($10^{-17}$~ergs~cm$^{-2}$~s$^{-1}$)}
&\colhead{\phantom{$>$}EW (\AA)} }
\startdata
10&4 01 32.46 &$-39$ 36 21.2 &25.60 &24.68 &18.4 &$\phantom{>}310 \pm 51$ \\
11 &4 01 27.66 &$-$39 44 06.6 &26.45 &\dots &8.4  &$> 82 \pm 15$ \\
12 &4 01 30.45 &$-$39 37 05.3 &26.89 &\dots &5.6  &$> 236 \pm 76$ \\
13 &4 01 29.55 &$-39$ 42 50.1 &26.94 &25.19 &5.3  &$\phantom{>}122 \pm 37$ \\
14 &4 02 40.71 &$-40$ 07 56.0 &25.46 &24.67 &20.9 &$\phantom{>}354 \pm 53$ \\
15 &4 02 34.61 &$-40$ 03 43.7 &26.35 &24.62 &9.2  &$\phantom{>}115 \pm 21$ \\
16 &4 02 24.04 &$-40$ 07 35.7 &26.54 &25.15 &7.7  &$\phantom{>}149 \pm 34$ \\
17 &4 03 04.49 &$-40$ 01 48.2 &26.92 &\dots &5.4  &$> 101 \pm 30$ \\
18 &4 01 31.68 &$-40$ 07 45.0 &25.02 &24.34 &31.3 &$\phantom{>}401 \pm 37$ \\
19 &4 01 21.03 &$-40$ 07 15.8 &26.12 &24.50 &11.4 &$\phantom{>}143 \pm 18$ \\
20 &4 00 31.02 &$-39$ 55 43.8 &26.19 &25.83 &10.7 &$\phantom{>}551 \pm 150$ \\
21 &4 00 37.25 &$-40$ 02 28.5 &26.95 &\dots &5.3  &$> 122 \pm 33$ \\
\enddata
\end{deluxetable}
\end{document}